\documentclass[12pt, english, a4paper]{amsart}
\usepackage{graphicx}
\usepackage{amsmath,amssymb}
\usepackage{bm}
\usepackage{euscript,color}
\usepackage{hyperref}
\usepackage{babel}
\usepackage{amsthm}
\usepackage{amsfonts}
\usepackage{latexsym}
\usepackage{fancyhdr}
\usepackage{enumerate}
\usepackage{multirow}
\usepackage{tabularx}
\usepackage{array}
 \usepackage{setspace}

\usepackage{ulem}

\usepackage{csquotes}
\usepackage[style=apa]{biblatex}
\usepackage{tikz}\usepackage{pgfplots}
\usetikzlibrary{intersections,calc}
\usetikzlibrary{arrows,decorations.markings}
\pgfplotsset{compat=1.12}
\tikzstyle arrowstyle=[scale=12pt]

\usepackage[left=2.5cm, top=3cm, bottom=3cm, right=2.5cm]{geometry}

\makeatletter
\renewcommand{\section}{\@startsection%
{section}
{1}
{0mm}
{2.5\bigskipamount}
{2.5\bigskipamount}
{\centering\normalsize\sc}}

\renewcommand{\paragraph}{\@startsection%
{paragraph}
{4}
{0mm}
{\bigskipamount}
{-1.25ex}
{\normalsize\sl}}

\def\provedboxcontents#1{$\square$}

\makeatother

\newtheoremstyle{thm}{}{}{\slshape}{}{\scshape}{.}{0.5em}{}

\newtheoremstyle{def}{}{}{}{}{\scshape}{.}{0.5em}{}

\newtheoremstyle{rmk}{}{}{}{}{\scshape}{.}{0.5em}{}

\newtheoremstyle{claim}{}{}{}{}{\slshape}{.}{0.5em}{}

\theoremstyle{thm}
\newtheorem{newstatement}{newstatement}

\newtheorem*{conjecture*}{Conjecture}

\theoremstyle{def}

\theoremstyle{rmk}

\newtheorem{example}[newstatement]{Example}

\theoremstyle{claim}

\expandafter\let\expandafter\oldproof\csname\string\proof\endcsname
\let\oldendproof\endproof
\renewenvironment{proof}[1][\proofname]{%
  \oldproof[\slshape #1]%
}{\oldendproof}

\let\epsilon\varepsilon

\renewcommand{\emph}[1]{{\slshape #1}}

\title[Coefficients of variation]{{  Comparison of} two coefficients of variation: a new Bayesian approach}

\author[F. Bertolino]{Francesco Bertolino}
  \email{bertolin@unica.it}

\author[S. Columbu]{Silvia Columbu*}
 \email{silvia.columbu@unica.it}

\author[M. Manca]{Mara Manca*}
 \email{mara.manca@unica.it}

\author[M. Musio]{Monica Musio*}
 \email{mmusio@unica.it}

 \address{*Dipartimento di Matematica e Informatica, Universit\`a degli Studi di
 Ca\-gli\-ari, Via Ospedale 72, 09124 Cagliari, Italy}

\date{}


\thanks{Research partially supported by the Fondazione di Sardegna 2020 research project ``Statistical models for the study of the socio-economic situations in Sardinia induced by the Covid-19 pandemic. Applications and comparisons with other Italian regions"}

\begin{document}

\begin{abstract}
The coefficient of variation is a useful indicator for comparing the spread of values between dataset with different units or widely different means. In this paper we address the problem  of investigating the equality of the coefficients of variation from two independent populations.  In order to do this we rely on the Bayesian Discrepancy Measure recently introduced in the literature. Computing this Bayesian measure of evidence is straightforward when the coefficient of variation is a function of a single parameter of the distribution. In contrast, it becomes difficult when it is a function of more parameters, often requiring the use of MCMC methods. We calculate the Bayesian Discrepancy Measure by considering a variety of distributions whose coefficients of variation depend on more than one parameter. We consider also applications to real data. As far as we know, some of the examined problems have not yet been covered in the literature.
\end{abstract}
\maketitle
\section{Introduction}

In many areas of applied statistics including medicine, biostatistics, anthropology, financial analysis, quality control, and chemical experiments, the coefficient of variation (CV) is commonly used as a measure of dispersion and it is applied to compare the relative variability of two or more independent populations. The problem of making inferences on the equality of two CVs has been widely addressed in the literature from a frequentist perspective following the Neyman-Pearson-Wald testing procedure. The majority of statistical approaches proposed to assess this equality are based on parametric models for hypothesis testing{  ,} and require the determination of the asymptotic distribution of the sample CV, which is often not easy to {  derive}. 
In the case of Normal populations one of the most recent contributions is given in \cite{krishnamoorthy2014improved}, where can also be found a quite exaustive review of the related literature. Many of the proposals discussed  in \cite{krishnamoorthy2014improved} are based on likelihood ratio statistics and their asymptotic properties. Various modified versions of the standard likelihood ratio statistic have also been discussed, such as the signed likelihood ratio (SLR) and a modified signed likelihood ratio  test (M-SLRT). The different testing procedures in the literature consider either the difference or the ratio of the CVs. {  It is usually} assumed that the data follows symmetric distributions, although asymptotic tests for asymmetric distributions have also been discussed. For instance, the case of the lognormal distribution is considered in \cite{nam2017inference} and \cite{Wong2019}. {  A more} general assumption of asymmetry, with applications to the Beta and the Gamma distributions, has been discussed in \cite{yue2019inference}. 
Nonparametric techniques, such as bootstrap, which free up the analysis from the parametric restrictive assumptions, have also alternatively been proposed (see \cite{cabras2006nonparametric} and \cite{amiri2010improvement}). Although being more flexible in terms of model assumptions, such approaches still rely on the definition of an appropriate test statistic.

{  There are little contributions to this problem in 
the Bayesian framework} and, as far as we know, they only refer to the case of Normal distributions. In particular \cite{lee2003bayesian} proposed an approach based on the use of fractional Bayes factors, whereas Pereira and Stern \cite{Pereira2001FullBS} proposed a measure of evidence for sharp null hypotheses called $e$-value and a related test.

All the literature cited above follows the decision-making framework for hypothesis testing, which consists of testing a null hypothesis against an alternative one.

In this paper, we instead propose to use a new Bayesian measure of evidence, called Bayesian Discrepancy Measure (BDM). We aim to assess the compatibility of the hypothesis on the equality of two coefficients of variation (CVs), from independent populations, with the observed data (see \cite{bertolino2021new}).

The proposed procedure shares the same idea of the Fisherian pure significance test (\cite{Fisher1925}){ ,} in which only a single hypothesis is specified (no alternative is considered) and where one checks the conformity or compatibility of the hypothesis with the observed data through a suitable evidence measure. Indeed, they are both based on the principle that ``\textit{Every experiment may be said to exist only in order to give the facts a chance of disproving the null hypothesis}''  (\cite{Fisher1925}). For a comparison of different approaches to statistical testing see \cite{Christensen2005}.


To highlight the simplicity and the potential of this method, different cases and models have been considered and examples on real data are presented. 

The structure of the paper is as follows. Section 2 provides a summary on the definition of the BDM and presents a general procedure to be followed {  for comparing a function of parameters in two independent populations}. Then, Section 3 deals with the comparison between two coefficients of variation and provides, on a case-by-case basis, the tools needed for that particular scenario, each followed by an application to real data, some of which are taken from the literature. Finally, the last section contains conclusions and guidelines to be followed for further research. As far as we know, some of the cases analysed have not been investigated before in the literature, see Section 3.3 concerning the Skew-Normal case and Section 3.4 for the Negative Binomial one.

\section{The Bayesian Discrepancy Test}
\label{Sec:2}
Let $X\sim f(x \vert \theta)$ be a parametric model indexed by a scalar parameter $\theta \in \Theta \subseteq \mathbb{R}$, $g_0(\theta)$ a prior density and $\boldsymbol{x}$ a sample of \textit{iid} observations. We denote by $g_1(\theta \vert \boldsymbol{x})\propto g_0(\theta)\; L(\theta \vert \boldsymbol{x})$ the posterior density, where $L(\theta \vert \boldsymbol{x})=\prod_{i=1}^{n} f(x_i \vert \theta)$ is the likelihood function. Let also $G_1(\theta \vert \boldsymbol{x})$ be the correspondent posterior distribution function, that we suppose strictly monotone,  and $m_1 = G_1^{-1} \big(\frac{1}{2} \big\vert \boldsymbol{x}\big)$ the posterior median. 

We are interested in the precise hypothesis $H:\theta=\theta_H$. Note that a precise, or sharp, hypothesis is a hypothesis consisting in a zero Lebesgue measure subset of the parameter space. The  BDM is a new measure of statistical evidence which is based on the evaluation of the discrepancy between the posterior median $m_1$ and  $\theta_H$. This naturally leads to consider the intervals $I_H$ (discrepancy interval) and $I_E$ (external interval) defined, in case of $\Theta=\mathbb{R}$, as

\begin{eqnarray}
I_H = \left\{
\begin{array}{ll}
(m_1,\theta_H) & \text{if} \quad  m_1 < \theta_H \\
\{m_1\} & \text{if} \quad  m_1 = \theta_H, \\
(\theta_H,m_1) & \text{if} \quad  m_1 > \theta_H \\
\end{array}
\right.
\end{eqnarray}

\begin{eqnarray}
I_E = \left\{
\begin{array}{ll}
(\theta_H,+\infty) & \text{if} \quad  m_1 < \theta_H \\
(-\infty,\theta_H) & \text{if} \quad   \theta_H < m_1. \\
\end{array}
\right.
\end{eqnarray}
When $m_1=\theta_H$, the external interval $I_E$ can be either $(-\infty, m_1)$ or $(m_1,+\infty)$.
Note that $\mathbb{P}(I_H \cup I_E)=\frac{1}{2}$. 
If  the support of the posterior is a subset of $\mathbb{R}$, the intervals $I_H$ and $I_E$ will be defined consequently. The BDM  is  then defined as
\begin{equation}
\label{formula:4}
\delta_H = 2 \cdot  \mathbb{P} (\theta \in I_H \vert \boldsymbol{x})=  1-2 \cdot  \mathbb{P} (\theta \in I_E \vert \boldsymbol{x}) = 1 - 2 \cdot \int_{I_E} d G_1(\theta \vert \boldsymbol{x})
\end{equation}
Note that $\delta_H $ can also be written as 
\begin{equation}
\label{formula:5}
\delta_H =1  -    2 \cdot \min
   \big\{ {\mathbb{P}}(\theta > \theta_H \vert \boldsymbol{x}) \, , \,
         {\mathbb{P}}(\theta < \theta_H \vert \boldsymbol{x}) \big\}.
\end{equation}        
This last formulation has the advantage of avoiding the computation of the posterior median. The more $\theta_H$ is far from the median of the posterior distribution $G_1(\theta \vert \boldsymbol{x}) $, the more $\delta_H$ is large and, in that case, $H$ is not conform to $G_1$. On the contrary, the smaller $\delta_H$ the stronger is the evidence in favour of $H$. Whether or not the data are compatible with the hypothesis depends of $\delta_H$ and it is a choice of the researcher that can eventually compare the results obtained with a given threshold.

Notice that the BDM is invariant  under invertible monotonic reparametrizations. Furthermore,  let $\theta^*$ indicate the true value of the parameter $\theta$, under general regularity conditions of identifiability, differentiability and integration (see \cite{cox1974distribution} and \cite{bertolino2021new}) and assuming the Cromwell’s Rule for the choice of the prior (see, for instance, \cite{lindley1991}), $\delta_H$ has the following properties: 
\begin{enumerate}
    \item [(i)] if $\theta_H = \theta^*$ then $\delta_H \sim Unif(\cdot\vert0, 1)$, for all sample sizes $n$;
    \item [(ii)] if $\theta_H \neq \theta^*$ then the consistency property holds, i.e. $\delta_H \xrightarrow{p} 1$.
\end{enumerate}

The approach described previously can be extended to the more general case of a parameter vector. Let $X\sim f(x \vert \boldsymbol{\theta})$, where $\boldsymbol{\theta}=(\theta_1,\ldots,\theta_k) \in \Theta \in \mathbb{R}^k$, $k>1$, and $\varphi=\gamma(\boldsymbol{\theta})$ be the parameter of interest, where $\gamma: \boldsymbol{\Theta} \to \Phi \subseteq \mathbb{R}$ is a known continuous function and there exists a bijective transformation from $\boldsymbol{\theta}$ to $(\varphi, \boldsymbol{\zeta})$, where $\boldsymbol{\zeta}$  is a nuisance parameter. It can happen that $\gamma(\boldsymbol{\theta}) = \theta_j$, for some $j$. Let also
\begin{equation}\label{ipotesiMULT}
 H: \varphi = \varphi_H
\end{equation}
be the precise hypothesis that we want to evaluate. Since the hypothesis induces a partition $\{\boldsymbol{\Theta}_a,\boldsymbol{\Theta}_H,\boldsymbol{\Theta}_b\}$ of the parameter space $\boldsymbol{\Theta}$, with

\begin{equation*}
    \begin{array}{ll}
        \boldsymbol{\Theta}_a\; &= \{\boldsymbol{\theta} \in \boldsymbol{\Theta} \;\vert \ \gamma(\boldsymbol{\theta}) < \varphi_H\}, \vspace{0.3cm}\\
        \boldsymbol{\Theta}_H &= \{\boldsymbol{\theta} \in \boldsymbol{\Theta} \;\vert \ \gamma(\boldsymbol{\theta}) = \varphi_H\}, \vspace{0.3cm}\\
        \boldsymbol{\Theta}_b\; &= \{\boldsymbol{\theta} \in \boldsymbol{\Theta} \;\vert\ \gamma(\boldsymbol{\theta}) > \varphi_H\},
    \end{array}
\end{equation*}

the definition of the BDM for the higher dimensional case is
 \begin{equation}\label{misuraDELTA-04}
   \begin{array}{ll} \vspace{4 pt}
   \delta_H   &= \; \displaystyle   1  -    2 \cdot \int_{I_E} {\rm d}G_1(\boldsymbol{\theta} \vert \boldsymbol{x}) \\ \vspace{3 pt}
   &= \displaystyle 1  -    2 \cdot \min_{a,b}
   \big\{ {\mathbb{P}}(\boldsymbol{\theta} \in \boldsymbol{\Theta}_a \ \vert \ \boldsymbol{x}) \, , \,
         {\mathbb{P}}(\boldsymbol{\theta} \in \boldsymbol{\Theta}_b \ \vert \ \boldsymbol{x}) \big\}     \, ,
   \,
 \end{array}
 \end{equation}
 where  now the external set $I_E$  is given by

\begin{equation}\label{insiemeEST}
  I_E \;  = \;
  \arg \min_{a,b}
  \big\{ {\mathbb{P}}(\boldsymbol{\theta} \in \boldsymbol{\Theta}_a \ \vert \ \boldsymbol{x}) \, , \,
         {\mathbb{P}}(\boldsymbol{\theta} \in \boldsymbol{\Theta}_b \ \vert\  \boldsymbol{x}) \big\}
         \;   .  \;
  \end{equation} 
If it is possible to compute the marginal posterior of $\varphi$ in a closed form, i.e. 
$$g_1^{M}(\varphi \mid \boldsymbol{x}) =\int_{\gamma(\boldsymbol{\theta})=\varphi} g_1(\boldsymbol{\theta} \mid \boldsymbol{x})\; d\boldsymbol{\theta},\quad \forall \varphi \in \Phi,$$
the BDM of the hypothesis \eqref{ipotesiMULT} is reconducted to the univariate case, see formula  (\ref{formula:4}). Methodological aspects of the BDM and similarities with Pereira and Stern's \cite{de1999evidence} e-value are discussed in \cite{bertolino2021new}.

The BDM is suitable for comparing parameter functions of two independent populations
$ X_\ell \sim f(\cdot \vert \boldsymbol{\theta}_\ell), \; \ell=1,2$. Given a prior density $g_0^\ell(\boldsymbol{\theta_\ell})$ and $\boldsymbol{x}_\ell=\{x_{\ell1}, \ldots, x_{\ell {n_{\ell}}} \}$ a sample of $n_\ell$ \textit{iid} observations, the posterior density is then $g_1^\ell(\boldsymbol{\theta_\ell} \vert \boldsymbol{x}_\ell) \propto g_0^\ell(\boldsymbol{\theta}_\ell)\; L^\ell(\boldsymbol{\theta}_\ell \vert \boldsymbol{x}_\ell)$. 
We indicate with $\boldsymbol{\theta}$ the joint parameter vector $\boldsymbol{\theta}=(\boldsymbol{\theta}_1,\boldsymbol{\theta}_2) \in \boldsymbol{\Theta}_1 \times \boldsymbol{\Theta}_2 = \boldsymbol{\Theta}$ of the two populations parameters.  We have that $\boldsymbol{\Theta}_\ell \subseteq \mathbb{R}^k$, where $k$ is the number of parameters in $\boldsymbol{\theta}_\ell$ and $\ell=1,2$.  
Consequently, $\boldsymbol{\Theta} \subseteq \mathbb{R}^{k}\times \mathbb{R}^k$.

We are interested to the hypothesis
\begin{equation}
\label{test}
    H:\varphi_1 - \varphi_2 = 0,
\end{equation}
which identifies the partition $\big\{ \boldsymbol{\Theta}_a, \, \boldsymbol{\Theta}_H, \, \boldsymbol{\Theta}_b \big\}$ of $\boldsymbol{\Theta}$, where

\begin{equation*}
    \begin{array}{ll}
        {\boldsymbol{\Theta}}_a &= \big\{ \boldsymbol{\theta} \in \boldsymbol{\Theta} \ \vert \ \varphi < \varphi_H \big\}, \vspace{0.3cm}\\
        {\boldsymbol{\Theta}}_H &= \big\{ \boldsymbol{\theta} \in \boldsymbol{\Theta} \ \vert \ \varphi = \varphi_H \big\}, \vspace{0.3cm}\\
        {\boldsymbol{\Theta}}_b &= \big\{ \boldsymbol{\theta} \in \boldsymbol{\Theta} \ \vert \ \varphi > \varphi_H \big\}.
    \end{array}
\end{equation*}
To compute the BDM (as seen in \eqref{misuraDELTA-04}) we need the   the minimum probability between ${\mathbb{P}}(\boldsymbol{\Theta}_a \ \vert \ \boldsymbol{x}_1, \boldsymbol{x}_2)$ and ${\mathbb{P}}(\boldsymbol{\Theta}_b \ \vert \ \boldsymbol{x}_1, \boldsymbol{x}_2 )$, where
\begin{equation}
    \mathbb{P}( \boldsymbol{\theta} \in \boldsymbol{\Theta}_j  \ \vert\ \boldsymbol{x}_1, \boldsymbol{x}_2)=\int_{\boldsymbol{\Theta}_j }  g^1_{1}(\boldsymbol{\theta}_1 \vert \boldsymbol{x}_1) \; g^2_{1}(\boldsymbol{\theta}_2 \vert \boldsymbol{x}_2) \; \mbox{d}\boldsymbol{\theta}_1\;\mbox{d}\boldsymbol{\theta}_2, \quad j=a,b.
\end{equation}
The evaluation of these probabilities require{  s} the computation of multidimensional integrals that can not be solved in a closed form, therefore we propose to approximate them using the Monte Carlo Integration method (the implemented \texttt{R} code is available in \cite{mara_manca_2022_7243897}). 

Since the distribution of $\xi=\varphi_1 - \varphi_2$ is univariate, in the event that it is also unimodal, the consistency property holds in analogy with what stated in Section \eqref{Sec:2}. Note that the same conclusions would also be reached by considering the hypothesis in the form $H: \frac{\varphi_1 }{\varphi_2 } = 1$, $\log \frac{\varphi_1 }{\varphi_2 }=0$, etc.


\section{Evaluating the equality of two CVs}
\label{Sec:3}

The CV is an adimensional parameter defined as the ratio of the standard deviation to the absolute value of the expected value
$$\varphi_\ell= \frac{\sqrt{Var(X_\ell)}}{ \vert E(X_\ell) \vert }, \quad \ell=1,2.$$
In some populations with several parameters it may happen that the CV depends only on one of them as, for instance, the case of a $LogN(\cdot\vert \mu,\sigma^2)$ model where $\text{CV}=\sqrt{e^{\sigma^2}-1}$, but also for some other populations such as $Gamma(\cdot \vert \alpha,\lambda)$, $Pareto(\cdot \vert \alpha,\lambda)$ and $Weibull(\cdot \vert\alpha,\lambda)$ where the CV depends solely on the shape parameter. 

In what follows, we propose the analysis of several non-trivial cases in which the comparison involves the full parameter space. In order to assess the consistency, the unimodality of the distribution of $\xi=\varphi_1 - \varphi_2$ has been verified numerically in all the cases considered.

In all of the examples we have adopted a Jeffrey's prior (see \cite{yang1996catalog} for a catalog of non-informative priors). However, other objective priors and, in the presence of substantive prior information, informative priors could equally be used.

\subsection{The case of two independent Normal populations}

Consider two indipendent Gaussian random variables $X_\ell \sim N(x_\ell \vert \mu_\ell,\phi_\ell^{-1})$,  with $X_\ell \in \mathbb{R}$ and mean and precision $(\mu_\ell,\phi_\ell) \in \mathbb{R} \times \mathbb{R}^+$, for $\ell=1,2$. Assuming the non-informative Jeffrey's priors 
$$(\mu_\ell,\phi_\ell)\sim g_0^{\ell} (\mu_\ell,\phi_\ell) \propto \phi_\ell^{-1}, \quad \ell=1,2,$$ 
and given $n_\ell$ observations with sample means $\Bar{x}_\ell$  and sample standard deviations $s_\ell$, it is known that the posterior distributions  of the parameter vectors are Normal Gamma 
$$(\mu_\ell,\phi_\ell)\ \vert \ \boldsymbol{x} \sim NG(\mu_\ell,\phi_\ell \vert \eta_\ell, \nu_\ell, \alpha_\ell, \beta_\ell), \quad \ell=1,2,$$   
with hyperparameters $\eta_\ell=\Bar{x}_\ell$, $\nu_\ell=n_\ell$, $\alpha_\ell=\frac{1}{2}(n_\ell-1)$, $\beta_\ell=\frac{1}{2}n_\ell s_\ell^2$. 

The hypothesis $H: \varphi_1-\varphi_2=0$, where $\varphi_\ell=\frac{1}{\vert\mu_\ell \vert \sqrt{\phi_\ell}}$, identifies in the parameter space $\boldsymbol{\Theta}$ the
subsets

\begin{equation*}
    \begin{array}{ll}
        \boldsymbol{\Theta}_a &=\big\{(\mu_1,\phi_1,\mu_2,\phi_2)\in \mathbb{R}^2\times \mathbb{R}^2_+ \ \big \vert \ \vert \mu_1 \vert \sqrt{\phi_1}> \vert \mu_2 \vert \sqrt{\phi_2}\big\}, \vspace{0.3cm} \\  
        \boldsymbol{\Theta}_H &=\big\{(\mu_1,\phi_1,\mu_2,\phi_2)\in \mathbb{R}^2\times \mathbb{R}^2_+ \ \big \vert \ \vert \mu_1 \vert \sqrt{\phi_1}= \vert \mu_2 \vert \sqrt{\phi_2}\big\},\vspace{0.3cm} \\ 
        \boldsymbol{\Theta}_b &=\big\{(\mu_1,\phi_1,\mu_2,\phi_2)\in \mathbb{R}^2\times \mathbb{R}^2_+ \ \big \vert \ \vert \mu_1 \vert \sqrt{\phi_1}< \vert \mu_2 \vert \sqrt{\phi_2}\big\}.
    \end{array}
\end{equation*}
The BDM (as seen in \eqref{misuraDELTA-04}) requires the computation of
\begin{equation*}
    \mathbb{P}\Big((\mu_1,\phi_1,\mu_2,\phi_2)\in \boldsymbol{\Theta}_j \ \Big \vert \ \boldsymbol{x} \Big)=\int_{\boldsymbol{\Theta}_j} \prod_{\ell=1}^2 g_1^\ell(\mu_\ell,\phi_\ell  \vert \ \eta_\ell, \nu_\ell, \alpha_\ell, \beta_\ell)\ \mbox{d}\mu_\ell \; \mbox{d}\phi_\ell, 
\end{equation*}
where $j=a,b$ and $g_1^\ell(\cdot)$ is the Normal Gamma density. For this setting we present a small simulation study to assess the proportion of erroneous conclusions when applying the BDM for varying sample sizes. An application to real data is also described.

\subsubsection{Simulation study}
We conducted simulation studies {  to evaluate} the behaviour of $\delta_H$ in repeated {  finite} sampling when comparing the CVs of two independent populations under the same distributional assumptions. 
Let $\boldsymbol{x}_1 = (x_{11}, \ldots, x_{1n_1})$ and $\boldsymbol{x}_2 = (x_{21}, \ldots, x_{2n_2})$ be two iid samples of size, respectively, $n_1$ and $n_2$ both taken from the Normal distribution $X \sim N(x \vert \mu, \phi^{-1})$. We set $\mu=\mu_1=\mu_2=3$ and $\phi=\phi_1=\phi_2=1$ to have that $\varphi_1-\varphi_2=0$. We are interested in assessing the compatibility of the hypothesis of equal CVs under this scenario. {  We considered 50000 Monte Carlo experiments and reported in Table \ref{table:TableSim} the proportion of times in which, for a selected threshold, the data falsely disagree with the hypothesis. The BDM was computed by taking $10000$ posterior draws.} The results obtained over resamplings are in agreement with the nominal accuracy typical of the first type error in the frequentist tests. This result is not surprising given the choice of objective priors (see \cite{Bayarri2004} and \cite{Hartigan1966} for other choices of the prior ). {  For the sake of comparison, we have also conducted similar experiments in the frequentist setting (see Table \ref{table:TableSim2}) by considering the non-parametric test proposed in \cite{cabras2006nonparametric}. We set $500$ bootstrap resamplings ( \cite{amiri2010improvement}). Under the same distributional assumptions, the two approaches have similar finite sample performances.}


\begin{example}  
Anthropometric measures in Sardinian population.
\end{example}

We consider a set of anthropometric measures concerning the Sardinian population. The sample consists of 280 individuals of both sexes (140 males and 140 females) aged 20--25, that was collected between 1995--1998. We focus on the CVs comparisons among men and women. The same data were presented and analysed in \cite{marini2005dispersion}, where the non-parametric bootstrap test for the difference of CVs developed in \cite{cabras2006nonparametric} was applied to evaluate sexual dimorphism. Among the 20 measurements in the dataset, we considered a subsample of 10 which can be assumed to be normally distributed. In Table \ref{table:TableNorm} are reported the principal descriptive statistics together with the values of the discrepancy measure associated to each anthropometric dimension considered. Based on the BDM{  ,} we conclude that only one hypothesis is not conform to the data, that is the equality between the coefficients of variation for men's and women's skinfolds tricepts. The final conclusions go in the same direction as \cite{marini2005dispersion}.


\subsection{The case of two independent inverse Gaussian populations}
Given two indipendent Inverse Gaussian random variables   
$X_\ell \sim IG(x_\ell \ \vert \ \mu_\ell,\lambda_\ell)$ i.e. 
$$f(x_\ell \ \vert \ \mu_\ell,\lambda_\ell)=\sqrt{\frac{\lambda_\ell}{2\pi x_\ell^3}} \exp{\Bigg\{ -\frac{1}{2}\lambda_\ell \Bigg(\frac{x_\ell-\mu_\ell}{\mu_\ell \sqrt{x_\ell}}\Bigg)^2\Bigg\}},$$
with $X_\ell \in \mathbb{R}^+$ and $(\mu_\ell,\lambda_\ell) \in \mathbb{R}^+ \times \mathbb{R}^+$ for $\ell=1,2$.

Assuming the Jeffreys non-informative priors $$(\mu_\ell,\lambda_\ell) \sim g_0^{\ell}(\mu_\ell,\lambda_\ell) \propto \frac{1}{\sqrt{{\mu_\ell}^3 \lambda_\ell}}, \quad \ell=1,2$$ 
and given $n_\ell$ observations, the posterior distribution of the parameter vector is
$$g_1^\ell(\mu_\ell,\lambda_\ell \ \vert \ \boldsymbol{x}_\ell) \propto \sqrt{\frac{\lambda_\ell^{n_\ell-1}}{\mu_\ell^3}} \exp{\Bigg\{ -\frac{n_\ell\lambda_\ell}{2} \Bigg(\frac{\bar{x}_\ell}{\mu_\ell^2} - \frac{2}{\mu_\ell} +\frac{1}{a_\ell}} \Bigg) \Bigg\}, \quad \ell=1,2$$
where $\bar{x}_\ell$ and $a_\ell$ are the arithmetic and harmonic means respectively. \\
The hypothesis $H: \varphi_1-\varphi_2=0$, where $\varphi_\ell=\sqrt{\frac{\mu_\ell}{\lambda_\ell}}$, identifies on the parameter space $\boldsymbol{\Theta}$ the
subsets 

\begin{equation*}
    \begin{array}{ll}
        \boldsymbol{\Theta}_a &=\Big\{(\mu_1,\lambda_1,\mu_2,\lambda_2)\in {\mathbb{R}}^4_+ \ \vert \ \mu_1 \lambda_2<\mu_2 \lambda_1\Big\}, \vspace{0.3cm} \\  
       \boldsymbol{\Theta}_H &=\Big\{(\mu_1,\lambda_1,\mu_2,\lambda_2)\in {\mathbb{R}}^4_+ \ \Big \vert \ \mu_1 \lambda_2= \mu_2 \lambda_1\Big\},\vspace{0.3cm} \\ 
        \boldsymbol{\Theta}_b &=\Big\{(\mu_1,\lambda_1,\mu_2,\lambda_2)\in {\mathbb{R}}^4_+ \ \Big \vert \ \mu_1\lambda_2>\mu_2\lambda_1\Big\}.
    \end{array}
\end{equation*}

The evaluation of the probabilities 
\begin{equation}
\label{prob_IG}
\mathbb{P}\Big( (\mu_1,\lambda_1,\mu_2,\lambda_2) \in \boldsymbol{\Theta}_j \ \Big\vert \ \boldsymbol{x}_1, \boldsymbol{x}_2\Big) = \int_{\boldsymbol{\Theta}_j} \prod_{\ell=1}^2 g_1^\ell(\mu_\ell,\lambda_\ell \ \vert \ \boldsymbol{x}_\ell)\ \mbox{d}\mu_\ell \ \mbox{d} \lambda_\ell , \quad j=a,b
\end{equation}
allows us to compute the BDM.

It is worth {  noting} that, in this case, the measure of skewness is three times the coefficient of variation. Therefore, comparing two skewness indices or two coefficients of variation is equivalent.

\noindent
\begin{example} 
Hodgkin's disease plasma bradykininogen levels.
\end{example}

We consider a study on the Hodgkin's disease in which plasma bradykininogen levels were measured for two groups of patients with active and inactive Hodgkin's disease. The outcome variable is measured in micrograms of bradykininogen per milliliter of plasma. This dataset was analyzed in \cite{chhikara1989inverse} where it was shown that the data fit an inverse Gaussian distribution. In the  group of patients with active disease we have that $n_1=17$ and $\varphi_1=0.2889$  while in the second group, with inactive Hodgkin's disease, $n_2=28$ and $\varphi_2=0.2810$. 

The computation of $\delta_H$ is linked to the evaluation of two four-dimensional integrals (see  \eqref{prob_IG}). We approximate the integrals through Monte Carlo methods.  
Since the posterior distribution is known up to a normalizing constant, it is not possible to sample directly from it. We therefore applied the random walk Metropolis-Hasting algorithm using a bivariate Normal proposal distribution. In order to obtain a chain that converges to the target distribution, $10^6$ samples were employed and reduced to $2\cdot 10^5$ after considering a burn-in period and a chain thinning.

Since the resulting discrepancy measure  is $\delta_H=0.2532$, we do not have enough evidence to assess the conformity to the data of the hypothesis on the CVs equality. The same dataset has been recently analysed in the frequentist frame in \cite{chankham2020confidence}, where the authors have considered four different versions of confidence intervals for the difference of CVs. The methods used were the Generalized Confidence Interval, the Adjusted Generalized Confidence Interval (\cite{krishnamoorthy2008inferences}, \cite{ye2010inferences}), the Method of Variance Estimation Recovery (MOVER) (\cite{donner2012closed}) and bootstrap percentiles. Exception made for the bootstrap percentiles, the remaining methods are based on the determination of pivotal quantities specific to the distribution assumed and require asymptotic approximations. Our method leads to the same conclusions as \cite{chankham2020confidence}.


\subsection{The case of two independent Skew Normal populations}
Consider two independent Skew Normal random variables $X_\ell \sim SN(x_\ell \ \vert \ \mu_\ell, \sigma_\ell, \lambda_\ell)$, $\ell=1,2$, whose density is  
\begin{align*}
    f(x_\ell \vert \mu_\ell,\sigma_\ell,\lambda_\ell) &= \frac{2}{\sigma_\ell} \phi\left(\frac{x_\ell-\mu_\ell}{\sigma_\ell}\right) \Phi\left(\lambda_\ell \frac{x_\ell-\mu_\ell}{\sigma_\ell}\right) \\
    &=\frac{2}{\sigma_\ell\sqrt{2\pi}}e^{-\frac{(x_\ell-\mu_\ell)^2}{2\sigma_\ell^2}}  \int_{-\infty}^{\lambda_\ell \frac{x_\ell-\mu_\ell}{\sigma_\ell}} \frac{1}{\sqrt{2\pi}} e^{-\frac{t_\ell^2}{2}}\; \mbox{d}t_\ell ,
\end{align*}
where $X_\ell \in \mathbb{R}$ and the location, the scale and the shape parameters are $(\mu_\ell,\sigma_\ell,\lambda_\ell) \in \mathbb{R} \times \mathbb{R}^+ \times \mathbb{R}$, while $\phi$ is the standard Normal probability density function and $\Phi$ is its cumulative distribution function. Given $\displaystyle \delta_\ell =\frac{\lambda_\ell}{\sqrt{\lambda_\ell^2+1}}$, the expected value, variance and coefficient of variation can be expressed as
\begin{align*}
E(X_\ell) &=\mu_\ell+\sigma_\ell\delta_\ell \sqrt{\frac{2}{\pi}},\\
Var(X_\ell)&= \sigma_\ell^2\Bigg(1-\frac{2\delta_\ell^2}{\pi}\Bigg),\\
CV(X_\ell) = \varphi_\ell &= \frac{\sqrt{\sigma_\ell^2\Big(1-\frac{2\delta_\ell^2}{\pi}\Big)}}{\Bigg|\mu_\ell+\sigma_\ell\delta_\ell \sqrt{\frac{2}{\pi}}\Bigg|}.
\end{align*}
The Jeffreys' priors can be approximated as
$$(\mu_\ell,\sigma_\ell,\lambda_\ell) \sim g_0^\ell(\mu_\ell,\sigma_\ell,\lambda_\ell) \propto \frac{1}{\sigma_\ell} g_0^\ell(\lambda_\ell),$$ 
where  $g_0^\ell(\lambda_\ell)$, the prior distribution of the shape parameter, is a generalized t-Student distribution with location parameter $\mu_\ell=0$, scale parameter $\sigma_\ell=\frac{1}{2}\pi$ and degrees of freedom $\nu_\ell=\frac{1}{2}$ (see \cite{bayes2007bayesian}), i.e. 
$$g_0^\ell(\lambda_\ell) = \frac{1}{B\big(\frac{\nu_\ell}{2},\frac{1}{2}\big)}\sqrt{\frac{\sigma_\ell}{\nu_\ell}}\left(1+\sigma_\ell \frac{(\lambda_\ell-\mu_\ell)^2}{\nu_\ell}\right)^{-\frac{\nu_\ell+1}{2}}.$$
The posterior distributions of the parameter vectors are then
\begin{equation*}
    g_1^\ell(\mu_\ell,\sigma_\ell,\lambda_\ell \vert \boldsymbol{x}_\ell) \propto \frac{1}{\sigma_\ell^{n_\ell+1}}\; g_0^\ell(\lambda_\ell)\; \prod_{i=1}^{n_\ell} \phi\left(\frac{x_{i\ell}-\mu_\ell}{\sigma_\ell}\right) \Phi\left(\lambda_\ell \frac{x_{i_\ell}-\mu_\ell}{\sigma_\ell}\right). 
\end{equation*}
The hypothesis $H: \varphi_1-\varphi_2=0$ identifies in the parameter space $\boldsymbol{\Theta}$ the
subsets 

\begin{equation*}
    \begin{array}{ll}
        \boldsymbol{\Theta}_a &=\left\{(\mu_1, \sigma_1, \lambda_1, \mu_2, \sigma_2, \lambda_2) \in {\mathbb{R}^4}\times{\mathbb{R}}^2_+ \ \Big|\  \varphi_1<\varphi_2 \right\}, \vspace{0.3cm} \\  
       \boldsymbol{\Theta}_H &=\left\{(\mu_1, \sigma_1, \lambda_1, \mu_2, \sigma_2, \lambda_2) \in {\mathbb{R}^4}\times{\mathbb{R}}^2_+ \ \Big|\  \varphi_1=\varphi_2 \right\},\vspace{0.3cm} \\ 
       \boldsymbol{\Theta}_b &=\left\{(\mu_1, \sigma_1, \lambda_1, \mu_2, \sigma_2, \lambda_2) \in {\mathbb{R}}^4\times{\mathbb{R}}^2_+ \ \Big|\ \varphi_1>\varphi_2\right\}.
    \end{array}
\end{equation*}

To compute BDM  we  evaluate the probabilities 

\begin{equation}
\label{prob_SKN}
\mathbb{P}\Big( (\mu_1, \sigma_1, \lambda_1, \mu_2, \sigma_2, \lambda_2) \in \boldsymbol{\Theta}_j  \ \Big|\ \boldsymbol{x}_1, \boldsymbol{x}_2\Big), \quad j=a,b.
\end{equation}
\noindent
\begin{example} 
Differential expression in colon cancer genes.
\end{example}

In the framework of genetic studies, microRNA (miRNA) levels are often used as diagnostic tools for cancer detection. Comparisons between two disease conditions, such as the presence or absence of cancer cells, often result in determining the presence of differentially expressed genes. A gene is said differentially expressed if there is a difference in its expression levels (abundance of miRNA) between two conditions. Usually{  ,} such differences are investigated by applying pair comparison statistical tests which rely on the assumption of normality distribution for gene's expression, and different transformations and normalizations of the original variable are applied in order to guarantee such condition. Instead, in a recent study  \cite{hossain2015application} proposed to assume a Skew Normal distribution for the gene's expression, since it helps to overcome the presence of possible asymmmetries. Following these suggestions we considered a study on colon cancer, analysed in \cite{hossain2015application}, and compared the CVs of a selection of genes in healthy and tumor tissues. The aim of the original work was to determine which miRNAs were differentially expressed between the two sets of samples collected during surgery from 145 subjects (116 with colon cancer and 29 healthy). Data is available from the public genomics data repository Gene Expression Omnibus (GEO) and accessible through GEO Series accession number GSE18392. 
In this analysis we consider a small selection of 3 miRNAs among those which were found to be differentially expressed (see \cite{sarver2009human}). \\
Given that we only know the posterior distribution up to a normalising constant, the application of MCMC methods is necessary in order to calculate the probabilities \eqref{prob_SKN}. In particular, following the literature on Bayesian inference for a Skew-Normal distribution (see \cite{bayes2007bayesian}), a Gibbs-sampler has been used. Chains convergence was reached with $6\cdot 10^5$ samples, that were reduced to $3\cdot 10^4$ after considering a burn-in period and a chain thinning.
In Table \ref{table:TableSkewNorm} are reported the principal descriptive statistics and the  values of the discrepancy measure. 

\subsection{The case of two independent Negative Binomial populations}
Let us consider two discrete Negative Binomial populations.  It is known that, given  $X_\ell \vert \lambda_\ell \sim \mbox{Poiss}(x_\ell \ |\ \lambda_\ell)$ with $\lambda_\ell \sim \mbox{Gamma}(\lambda_\ell \ |\ \alpha_\ell,\beta_\ell)$, then the unconditional random variables $X_\ell$ follow a Negative Binomial distribution 

$$ X_\ell \sim NB\Big(x_\ell \ \Big\vert\ \frac{\beta_\ell}{\beta_\ell+1},\alpha_\ell\Big), \quad \ell=1,2$$

with $(\alpha_\ell,\beta_\ell) \in \mathbb{R}^+ \times \mathbb{R}^+$. Its expected value, variance and coefficient of variation are
\begin{align*}
E(X_\ell)&=\frac{\alpha_\ell}{\beta_\ell},\\
Var(X_\ell)&=\alpha_\ell \frac{\beta_\ell+1}{\beta_\ell^2},\\
CV(X_\ell)&= \varphi_\ell = \sqrt{\frac{\beta_\ell+1}{\alpha_\ell}}.
\end{align*}

\noindent {  Given the Jeffreys' priors for $\lambda_\ell$ (see \cite{yang1996catalog}) in the form
$$g_0^\ell(\alpha_\ell, \beta_\ell) \propto \frac{1}{\beta_\ell} \sqrt{\alpha_\ell \psi^{(1)}(\alpha_\ell)-1}, \quad \ell=1,2$$
then the posterior distributions of the parameter vectors take the expressions. }
\begin{equation*}
    g_1^\ell(\alpha_\ell,\beta_\ell \ \vert\ \boldsymbol{x}_\ell) \propto \frac{\prod_i(x_{i\ell}+\alpha_\ell-1)!}{\left[(\alpha_\ell-1)!\right]^{n_\ell}} 
    \left[\frac{\beta_\ell^{\alpha_\ell}}{(\beta_\ell+1)^{\alpha_\ell+\bar{x}_\ell}}\right]^n \frac{1}{\beta_\ell} \sqrt{\alpha_\ell \psi^{(1)}(\alpha_\ell)-1}.
\end{equation*}
{  Note that $\psi^{(1)}(\alpha_\ell)=\sum_{j=0}^{\infty}(\alpha_\ell+j)^{-2}$ is the PolyGamma function.}\\
The hypothesis $H: \varphi_1 - \varphi_2 = 0$ identifies on the parameter space $\boldsymbol{\Theta}$ the subsets

\begin{equation*}
    \begin{array}{ll}
        \boldsymbol{\Theta}_a &=\Big\{(\alpha_1, \beta_1, \alpha_2, \beta_2) \in \mathbb{R}^2_+ \times \mathbb{R}^2_+ \ \Big|\ \alpha_2(\beta_1+1)<\alpha_1(\beta_2+1)\Big\}, \vspace{0.3cm} \\  
       \boldsymbol{\Theta}_H &=\Big\{(\alpha_1, \beta_1, \alpha_2, \beta_2) \in \mathbb{R}^2_+ \times \mathbb{R}^2_+ \ \Big|\ \alpha_2(\beta_1+1) = \alpha_1(\beta_2+1)\Big\},\vspace{0.3cm} \\ 
       \boldsymbol{\Theta}_b &=\Big\{(\alpha_1, \beta_1, \alpha_2, \beta_2) \in \mathbb{R}^2_+ \times \mathbb{R}^2_+ \ \Big|\ \alpha_2(\beta_1+1)<\alpha_1(\beta_2+1) \Big\},
    \end{array}
\end{equation*}
for which we have to compute the probabilities 

\begin{equation}
\label{prob_NB}
\mathbb{P}\Big( (\alpha_1, \beta_1, \alpha_2, \beta_2) \in \boldsymbol{\Theta}_j  \ \Big|\ \boldsymbol{x}_1, \boldsymbol{x}_2 \Big), \quad j=a,b.
\end{equation}

\noindent
\begin{example} 
Superspreading of COVID-19 infections
\end{example}
In the epidemiology context, when investigating the spreading dynamics of an infectious disease, it is often of interest to model the number of new infections (second cases) generated by an infectious subject. This kind of variable, in cases of individual variability in the transmission patterns, is overdispersed and right skewed (see \cite{lloyd2005superspreading}), therefore the negative binomial distribution is often considered for its analysis. During the Sars-Cov-2 (COVID-19) pandemic many national health systems have collected such kind of information considering contact tracing data to {  identify the diffusion pathways} of the disease and the behaviours leading to {  contagion growth}. {  It was observed that a small number of infected could cause most of the secondary infections, therefore differences in individual contact patterns were a symptom of a superspreading event.} Looking at the recent literature we used two published {  datasets} on secondary cases of COVID-19 and compared their CVs through the BDM. The first dataset considered comes from data tracing in two Indian states (see \cite{laxminarayan2020epidemiology}) from March to July 2020. It is a large dataset containing the offspring distribution for 88,527 cases, with sample mean 0.48 and variance 1.15 resulting in a CV of 2.218. The second one contains 290 cases from Hong Kong (see \cite{adam2020clustering}) recorded from January to April 2020. In this case the observed mean is 0.58 whereas the variance 1.29 leading to a sample CV of 2.217. This two coefficients of variation are extremely close between each other and, as expected, we get to a small BDM of $\delta_H=0.00972$. 

\begin{figure}[ht]
\begin{center}
\includegraphics[width=12cm, height=8cm]{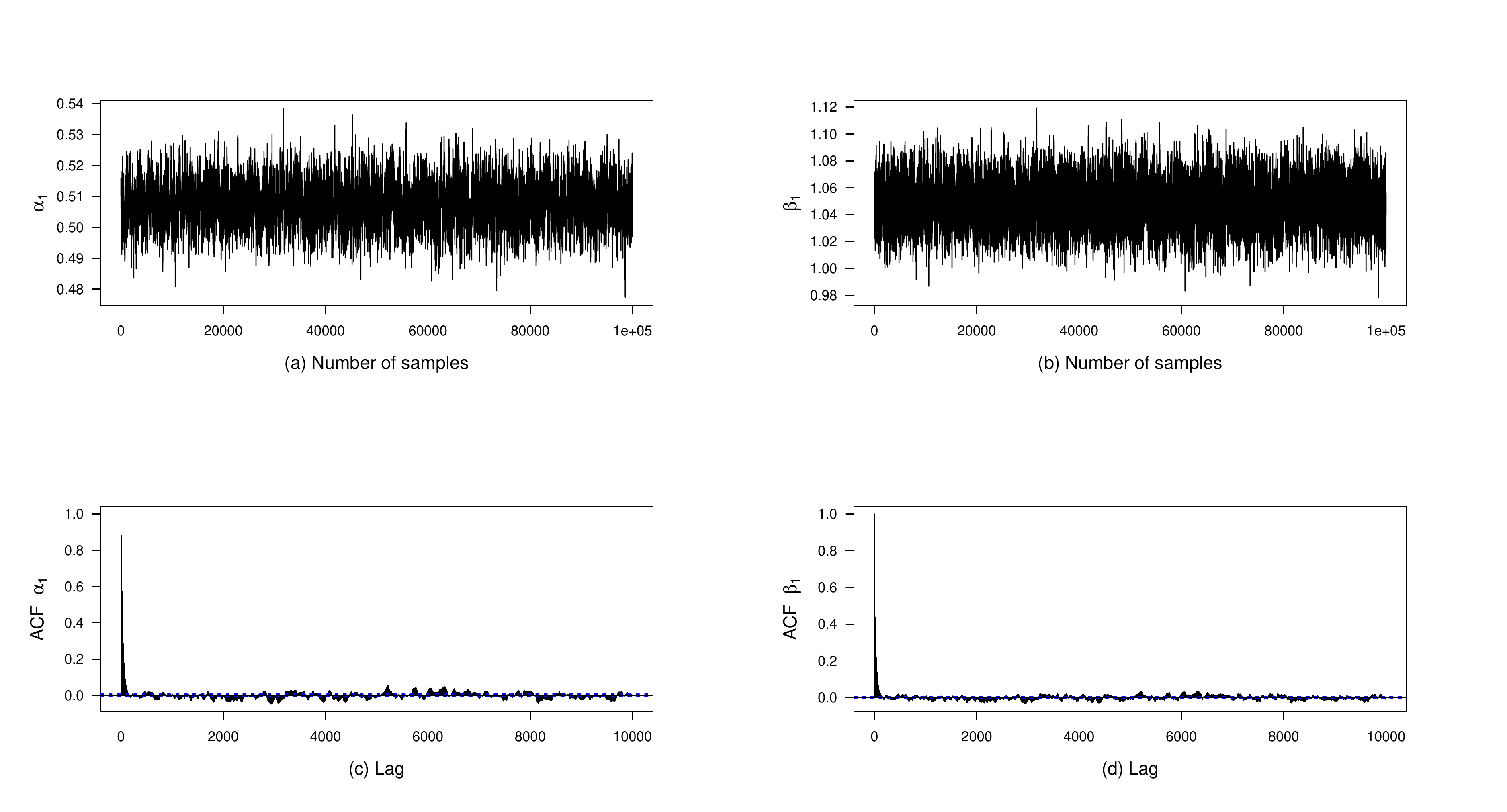}
\caption{Trace plots and autocorrelation functions for the MCMC chains of the Negative Binomial distribution parameters for the first population in example 4. Data coming from a tracing study on COVID-19 spread conducted in two Indian states.}\label{fig:convindia}
\end{center}\end{figure}

\begin{figure}[ht]
\begin{center}
\includegraphics[width=12cm, height=8cm]{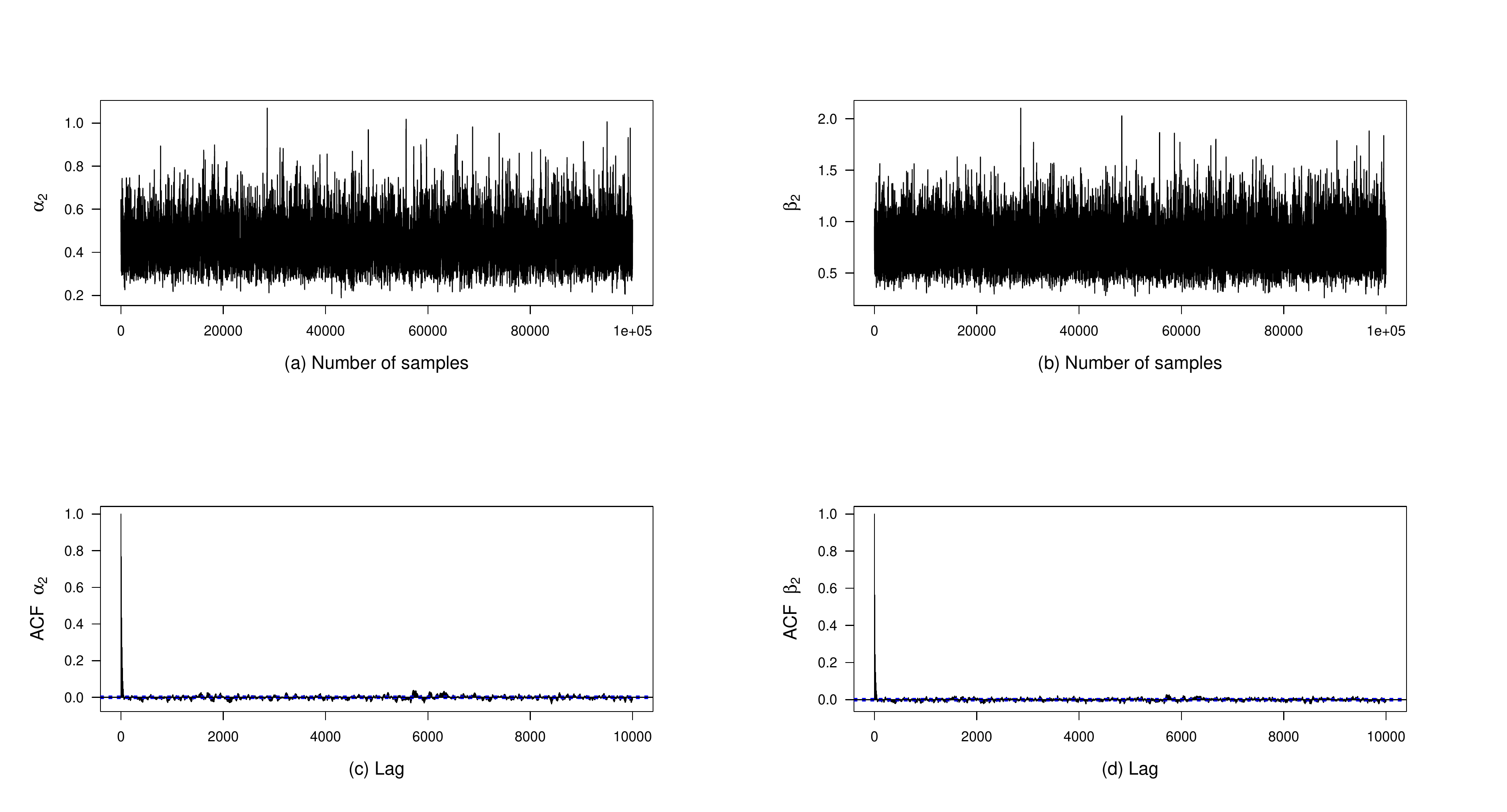}
\caption{Trace plots and autocorrelation functions for the MCMC chains of the Negative Binomial distribution parameters for the second population in example 4. Data coming from a study on COVID-19 spread conducted in Hong Kong.}\label{fig:convhongkong}
\end{center}\end{figure}

Also for this last model, the application of Metropolis-Hasting algorithms was required for the evaluation of probabilities \eqref{prob_NB}. In order to obtain a chain that converged to the target distribution, $1.3\cdot 10^5$ samples were employed and reduced to $10^5$ after considering a burn-in period, {  see Figures \ref{fig:convindia} and \ref{fig:convhongkong} for a graphical illustration. Diagnostics for the other examples in the paper can be found in \cite{mara_manca_2022_7243897}.}

\section{Discussion and conclusions}
 In this paper we have proposed the application of the BDM to assess if assuming that two independent populations have the same CV is conform or not with the data. The presented procedure is general and it does not require ``ad hoc'' techniques such as those proposed to solve similar problems in the frequentist settings, which are usually applicable only to particular cases. {  We have discussed four different applications, it would be interesting to explore also the application to new distributions such as those recently proposed in \cite{Ong2023}, \cite{Korkmaz2017}, \cite{MirMostafaee2017}.}

Furthermore, as far as we know, the cases of the Skew Normal populations and Negative Binomial ones are completely original and they have not been addressed before in the literature. 

The method outlined here has a more general validity as it can be extended to comparisons between parameters of two independent populations and/or their functions. This is feasible without making use of any asymptotic technique. Extensions include comparisons between skewness and kurtosis coefficients, regression coefficients, etc.


\nocite{*}
\printbibliography

@article{krishnamoorthy2014improved,
	author = {Krishnamoorthy, K. and Lee, M.},
	journal = {Computational Statistics},
	number = {1},
	pages = {215--232},
	publisher = {Springer},
	title = {Improved tests for the equality of normal coefficients of variation},
	volume = {29},
	year = {2014}}

@article{nam2017inference,
	author = {Nam, J. M. and Kwon, D.},
	journal = {Communications in Statistics-Theory and Methods},
	number = {17},
	pages = {8575--8587},
	publisher = {Taylor \& Francis},
	title = {Inference on the ratio of two coefficients of variation of two lognormal distributions},
	volume = {46},
	year = {2017}}

@article{Wong2019,
  author = {Wong, A. and Jiang, L.},
  journal = {Journal of Probability and Statistics},
  title = {Improved Small Sample Inference on the
Ratio of Two Coefficients of Variation of Two Independent Lognormal
Distributions},
  volume = {46},
  number = {17},
  pages = {8575--8587},
  year = {2019}
}

@article{yue2019inference,
	author = {Yue, Z. and Baleanu, D.},
	journal = {Symmetry},
	number = {6},
	doi = {824},
	publisher = {Multidisciplinary Digital Publishing Institute},
	title = {Inference about the Ratio of the Coefficients of Variation of Two Independent Symmetric or Asymmetric Populations},
	volume = {11},
	year = {2019}}

@article{cabras2006nonparametric,
	author = {Cabras, S. and Mostallino, G. and Racugno, W.},
	journal = {Communications in Statistics-Simulation and Computation},
	number = {3},
	pages = {715--726},
	publisher = {Taylor \& Francis},
	title = {A nonparametric bootstrap test for the equality of coefficients of variation},
	volume = {35},
	year = {2006}}

@article{amiri2010improvement,
	author = {Amiri, S. and Zwanzig, S.},
	journal = {Communications in Statistics: Simulation and Computation},
	number = {9},
	pages = {1726--1734},
	publisher = {Taylor \& Francis},
	title = {An improvement of the nonparametric bootstrap test for the comparison of the coefficient of variations},
	volume = {39},
	year = {2010}}

@article{lee2003bayesian,
	author = {Lee, H. C. and Kang, S. G. and Kim, D. H.},
	journal = {Journal of the Korean Data and Information Science Society},
	number = {4},
	pages = {1023--1030},
	publisher = {Korean Data and Information Science Society},
	title = {Bayesian test for equality of coefficients of variation in the normal distributions},
	volume = {14},
	year = {2003}}

@article{madruga2003bayesian,
	author = {Madruga, M. R. and Pereira, C.A.B. and Stern, J. M.},
	journal = {Journal of Statistical Planning and Inference},
	number = {2},
	pages = {185--198},
	publisher = {Elsevier},
	title = {Bayesian evidence test for precise hypotheses},
	volume = {117},
	year = {2003}}

@article{de1999evidence,
	author = {Pereira, C.A.B. and Stern, J. M.},
	journal = {Entropy},
	number = {4},
	pages = {99--110},
	publisher = {Molecular Diversity Preservation International},
	title = {Evidence and credibility: full Bayesian significance test for precise hypotheses},
	volume = {1},
	year = {1999}}

@inproceedings{Pereira2001FullBS,
  title={Full Bayesian Significance Test for Coefficients of Variation},
  author={Carlos Alberto De Bragança Pereira and Julio Michael Stern},
  year={2001}
}

@article{bertolino2021new,
	author = {Bertolino, F. and Manca, M. and Musio, M. and Racugno, W. and Ventura, L.},
	journal = {arXiv preprint arXiv:2105.13716v3},
	title = {A new Bayesian discrepancy measure},
	year = {2022}}

@article{marini2005dispersion,
	author = {Marini, E. and Rebato, E. and Racugno, W. and Buffa, R. and Salces, I. and Borgognini Tarli, S. M.},
	journal = {American Journal of Physical Anthropology: The Official Publication of the American Association of Physical Anthropologists},
	number = {3},
	pages = {342--350},
	publisher = {Wiley Online Library},
	title = {Dispersion dimorphism in human populations},
	volume = {127},
	year = {2005}}

@article{bayes2007bayesian,
	author = {Bayes, C. L. and Branco, M.D.},
	journal = {Brazilian Journal of Probability and Statistics},
	pages = {141--163},
	publisher = {JSTOR},
	title = {Bayesian inference for the skewness parameter of the scalar skew-normal distribution},
	year = {2007}}

@book{yang1996catalog,
	author = {Yang, R. and Berger, J. O.},
	publisher = {Institute of Statistics and Decision Sciences, Duke University},
	title = {A catalog of noninformative priors},
	year = {1996}}

@article{eilam1968bradykininogen,
	author = {Eilam, N. and Johnson, P. K. and Johnson, N. L. and Creger, W. P.},
	journal = {Cancer},
	number = {3},
	pages = {631--634},
	publisher = {Wiley Online Library},
	title = {Bradykininogen levels in Hodgkin's disease},
	volume = {22},
	year = {1968}}

@inproceedings{chankham2020confidence,
	author = {Chankham, W. and Niwitpong, S. A. and Niwitpong, S.},
	booktitle = {International Symposium on Integrated Uncertainty in Knowledge Modelling and Decision Making},
	organization = {Springer},
	pages = {372--383},
	title = {Confidence Intervals for the Difference Between the Coefficients of Variation of Inverse Gaussian Distributions},
	year = {2020}}

@book{chhikara1989inverse,
	author = {Chhikara, R.S. and Folks, J.L.},
	journal = {Methodology and Applications},
	pages = {7--20},
	publisher = {Marcel Dekker, New York},
	title = {The Inverse Gaussian Distribution: Theory},
	year = {1989}}

@article{hossain2015application,
	author = {Hossain, A. and Beyene, J.},
	journal = {Journal of Applied Statistics},
	number = {3},
	pages = {477--491},
	publisher = {Taylor \& Francis},
	title = {Application of skew-normal distribution for detecting differential expression to microRNA data},
	volume = {42},
	year = {2015}}

@article{laxminarayan2020epidemiology,
	author = {Laxminarayan, R. and Wahl, B. and Dudala, S. R. and Gopal, K. and Mohan B. C. and Neelima, S. and Jawahar Reddy, K. S. and Radhakrishnan, J. and Lewnard, J. A.},
	journal = {Science},
	number = {6517},
	pages = {691--697},
	publisher = {American Association for the Advancement of Science},
	title = {Epidemiology and transmission dynamics of COVID-19 in two Indian states},
	volume = {370},
	year = {2020}}

@article{adam2020clustering,
	author = {Adam, D. C. and Wu, P. and Wong, J. Y. and Lau, E. H. Y.and Tsang, T. K. and Cauchemez, S. and Leung, G. M. and Cowling, B. J.},
	journal = {Nature Medicine},
	number = {11},
	pages = {1714--1719},
	publisher = {Nature Publishing Group},
	title = {Clustering and superspreading potential of SARS-CoV-2 infections in Hong Kong},
	volume = {26},
	year = {2020}}

@article{lloyd2005superspreading,
	author = {Lloyd-Smith, J. O. and Schreiber, S. J. and Kopp, P. E. and Getz, W. M.},
	journal = {Nature},
	number = {7066},
	pages = {355--359},
	publisher = {Nature Publishing Group},
	title = {Superspreading and the effect of individual variation on disease emergence},
	volume = {438},
	year = {2005}}

@article{donner2012closed,
	author = {Donner, A. and Zou, G.Y.},
	journal = {Statistical Methods in Medical Research},
	number = {4},
	pages = {347--359},
	publisher = {Sage Publications Sage UK: London, England},
	title = {Closed-form confidence intervals for functions of the normal mean and standard deviation},
	volume = {21},
	year = {2012}}

@article{krishnamoorthy2008inferences,
	author = {Krishnamoorthy, K. and Tian, L.},
	journal = {Journal of Statistical Planning and Inference},
	number = {7},
	pages = {2082--2089},
	publisher = {Elsevier},
	title = {Inferences on the difference and ratio of the means of two inverse Gaussian distributions},
	volume = {138},
	year = {2008}}

@article{ye2010inferences,
	author = {Ye, R. D. and Ma, T. F. and Wang, S. G.},
	journal = {Computational statistics \& data analysis},
	number = {4},
	pages = {906--915},
	publisher = {Elsevier},
	title = {Inferences on the common mean of several inverse Gaussian populations},
	volume = {54},
	year = {2010}}

@article{sarver2009human,
  title={Human colon cancer profiles show differential microRNA expression depending on mismatch repair status and are characteristic of undifferentiated proliferative states},
  author={Sarver, A. L. and French, A. J. and Borralho, P. M. and Thayanithy, V. and Oberg, A. L. and Silverstein, K. A.T. and Morlan, B. W. and Riska, S. M. and Boardman, L. A. and Cunningham, J. M. and others},
  journal={BMC cancer},
  volume={9},
  number={1},
  pages={1--15},
  year={2009},
  publisher={BioMed Central}
}

@inbook{cox1974distribution,
  author = {Cox, D. R. and Hinkley, D. V. },
  title     = {Distribution-free and randomization tests},
  chapter   = {Theoretical Statistics},
  publisher = {Chapman and Hall/CRC Press},
  edition   = {1st ed.},
  year      = {1979},
  pages     = {179--206}
}

@book{lindley1991,
	author = {Lindley, D. V. },
	title = {Making Decisions},
	publisher = {John Wiley and Sons Inc},
	address   = {United States},
	edition   = {2nd ed.},
	year = {1991}
	}

@article{Christensen2005,
  author={Christensen, R.},
  title={Testing Fisher, Neyman, Pearson, and Bayes},
  journal={The American Statistician},
  volume={59},
  number={2},
  pages={121--126},
  year={2005}
}

@book{Fisher1925,
  author={Fisher, R.A.},
  title={Statistical Methods for Research Workers},
  publisher={Oliver and Boyde},
  volume={59},
  address={Edinburgh},
  year={1925}
}

@article{Bayarri2004,
  author={Bayarri, M.J. and Berger, J.O.},
  title={The Interplay of Bayesian and Frequentist Analysis},
  journal={Statistical Science},
  volume={19},
  number={1},
  pages={58--80},
  year={2004}
}

@software{mara_manca_2022_7243897,
  author       = {Mara Manca and
                  Silvia Columbu},
  title        = {{maramanca/CVs\_comparison: Testing the equality of 
                   two coefficients of variation: a new Bayesian
                   approach}},
  month        = oct,
  year         = 2022,
  publisher    = {Zenodo},
  version      = {v1.0.0},
  doi          = {10.5281/zenodo.7243897},
  url          = {https://doi.org/10.5281/zenodo.7243897}
}

@article{Hartigan1966,
  author={Hartigan, J.A.},
  title={Note on the Confidence-Prior of Welch and Peers},
  journal={Journal of the Royal Statistical
Society: Series B (Methodological)},
  volume={28},
  number={1},
  pages={55--56},
  year={1966}
}

@article{Ong2023,
  author={Ong, S.H. and Sim, S.Z.},
  title={A bivariate distribution with convolution of binomials as marginals},
  journal={Communications in Statistics: Simulation and Computation},
  doi = {10.1080/03610918.2023.2171060},
  year={2023}
}

@article{Korkmaz2017,
  author={Korkmaz, M.C.},
  title={A generalized skew slash distribution via gamma-normal distribution},
  journal={Communications in Statistics: Simulation and Computation},
  volume={46},
  pages={1647--1660},
  year={2017}
}

@article{MirMostafaee2017,
  author={MirMostafaee, S.M.T.K. and Mahdizadeh, M. and Lemonte, A.J.},
  title={The Marshall-Olkin extended generalized Rayleigh distribution: Properties and applications},
  journal={Communications in Statistics: Theory and Methods},
  volume={46},
  pages={653--671},
  year={2017}
}
\newpage
\begin{center}
\begin{table}[t]
\caption{False non-conformity rates (FNCR) for different sample sizes $n_1$ and $n_2$ and three different thresholds. Model $N(3,1)$. } 
\label{table:TableSim}
\resizebox{\columnwidth}{!}{
{\setlength{\extrarowheight}{1pt}
\begin{tabular}{|c|cccc|cccc|cccc|}
\hline
\textit{} & \multicolumn{4}{|c|}{$0.90$}   & \multicolumn{4}{|c|}{$0.95$} & \multicolumn{4}{|c|}{$0.99$}         \\
\hline
$n_1$ & 10 & 10 & 100 & 1000 &  10 & 10 & 100 & 1000 & 10 & 10 & 100 & 1000 \\
$n_2$ & 10 & 50 & 100 & 1000 &  10 & 50 & 100 & 1000 & 10 & 50 & 100 & 1000 \\ \hline
\hline
FNCR & 0.096 & 0.105 & 0.1 & 0.098 & 0.047 & 0.054 & 0.05 & 0.049 &0.009 & 0.011 & 0.01 & 0.01 \\
\hline
\end{tabular}}}
\end{table}
\end{center}

\begin{center}
\begin{table}[t]
\caption{False positive rates (FPR) for different sample sizes $n_1$ and $n_2$ and three different thresholds for the non-parametric bootstrap test in the frequentist frame. Model tested $N(3,1)$.}
\label{table:TableSim2}
\resizebox{\columnwidth}{!}{
{\setlength{\extrarowheight}{1pt}
\begin{tabular}{|c|cccc|cccc|cccc|}
\hline
\textit{} & \multicolumn{4}{|c|}{$0.10$}   & \multicolumn{4}{|c|}{$0.05$} & \multicolumn{4}{|c|}{$0.01$}         \\
\hline
$n_1$ & 10 & 10 & 100 & 1000 &  10 & 10 & 100 & 1000 & 10 & 10 & 100 & 1000 \\
$n_2$ & 10 & 50 & 100 & 1000 &  10 & 50 & 100 & 1000 & 10 & 50 & 100 & 1000 \\ \hline
\hline
FPR & 0.083 & 0.102 & 0.980 & 0.088 & 0.049 & 0.046 & 0.053 & 0.044 &0.008 & 0.012 & 0.0097 & 0.014 \\
\hline
\end{tabular}}}
\end{table}
\end{center}

\begin{center}
\begin{table}[t]
\caption{Dispersion dimorphism in a set of anthropometric dimensions and relative BDM. Weight is expressed in kg; skinfolds in mm; all other measurements in cm.}
\label{table:TableNorm}
\resizebox{\columnwidth}{!}{
{\setlength{\extrarowheight}{1pt}
\begin{tabular}{lccccccccccc}
\hline
\textit{}                 & \multicolumn{4}{c}{\textbf{Men}}            & \multicolumn{1}{l}{} & \multicolumn{4}{c}{\textbf{Women}}          & \multicolumn{1}{l}{} & \multirow{2}{*}{$\delta_H$} \\ \cline{2-5} \cline{7-10}
\textit{}                 & Mean  & SD   & $n_1$   & \multicolumn{1}{l}{CV} &                      & Mean  & SD   & $n_2$   & \multicolumn{1}{l}{CV} & \multicolumn{1}{l}{} &                             \\ \hline
\hspace{0.1mm} \textit{Weight}           & 67.22 & 8.46 & 140 & 0.126                  &                      & 53.71 & 7.59 & 140 & 0.141                  &                      & 0.812                       \\
 Breadths &  &  &  &   & & & & & &  &  \\
\hspace{0.1mm} \textit{Cephalic} & 15.10 & 0.64 & 141 & 0.042                  &                      & 14.53 & 0.58 & 172 & 0.040                   &                      & 0.550                       \\
\hspace{0.1mm} \textit{Elbow}            & 7.02  & 0.39 & 103 & 0.056                  &                      & 6.01  & 0.35 & 117 & 0.058                  &                      & 0.355                       \\
Circumferences &  &  &  &   & & & & & &  &  \\
\hspace{0.1mm} \textit{Midarm relaxed}   & 26.91 & 2.60 & 139 & 0.097                  &                      & 23.47 & 2.01 & 134 & 0.086                  &                      & 0.831                       \\
\hspace{0.1mm} \textit{Midarm tensed}    & 30.83 & 2.74 & 139 & 0.089                  &                      & 25.28 & 2.15 & 133 & 0.085                  &                      & 0.388                       \\
Skinfolds &  &  &  &   & & & & & &  &  \\

\hspace{0.1mm} \textit{Biceps}           & 4.10  & 1.79 & 137 & 0.437                   &                      & 6.08  & 2.51 & 133 & 0.413                   &                      & 0.420                       \\
\hspace{0.1mm} \textit{Triceps}          & 7.76  & 3.76 & 140 & 0.485                   &                      & 13.33 & 4.78 & 140 & 0.359                   &                      & \underline{0.996}                       \\
\hspace{0.1mm} \textit{Subscapular}      & 10.34 & 3.78 & 137 & 0.366                   &                      & 12.71 & 4.53 & 140 & 0.356                   &                      & 0.213                      \\
\hspace{0.1mm} \textit{Suprailiac}       & 9.23  & 4.34 & 140 & 0.470                   &                      & 10.21 & 4.48 & 140 & 0.439                   &                      & 0.507                      \\
\hspace{0.1mm} \textit{Abdominal}        & 12.15 & 6.52 & 97  & 0.537                   &                      & 12.77 & 5.74 & 111 & 0.449                   &                      & 0.848                      \\ \hline
\end{tabular}}}
\end{table}
\end{center}


\begin{center}
\begin{table}[t]
\caption{Three differentially expressed miRNAs in the colon cancer study.}
\label{table:TableSkewNorm}
\small
\begin{tabular}{lccccccccc}
\hline
\textit{}            & \multicolumn{3}{c}{\textbf{Tumor tissue}}            & \multicolumn{1}{l}{} & \multicolumn{3}{c}{\textbf{Healthy tissue}}          & \multicolumn{1}{l}{} & \multirow{2}{*}{$\delta_H$} \\ \cline{2-4} \cline{5-8}
\textit{}            & Mean  & SD      & \multicolumn{1}{l}{CV} &                      & Mean  & SD     & \multicolumn{1}{l}{CV} & \multicolumn{1}{l}{} &                             \\ \hline
\textit{miR-182} & 13.9 & 0.59 & 0.04 &                       & 13.32 & 0.32 & 0.02 &                      & 0.9997 \\
\textit{miR-183} & 13.17 &  0.72 & 0.05 &                     & 12.32 & 0.43 & 0.04 &                     & 0.9964 \\ 
\textit{miR-96}  & 10.75 & 0.85 & 0.08 &                      & 10.19 & 0.41  & 0.04 &                     & 1         \\
\hline
\end{tabular}
\end{table}
\end{center}

\end{document}